\documentclass[a4paper, 10pt]{article}
\usepackage{pslatex} % Font selection
\usepackage{chngpage}
\usepackage[tmargin=2.5cm,bmargin=2.5cm,lmargin=2.5cm,rmargin=2.5cm]{geometry} % Margins
\usepackage{graphicx} % Useful package for graphics inclusion
\usepackage{psfrag}
\usepackage{setspace}
\usepackage{float}

\usepackage{balance}
\usepackage{amssymb,amsmath}

\usepackage[compact]{titlesec}
\titleformat{\section}{\large\bfseries}{\thesection.~}{0pt}{}
\titlespacing*{\section}{0pt}{20pt}{10pt}
\titleformat{\subsection}{\normalfont\bfseries}{\thesubsection.~}{0pt}{}
\titlespacing*{\subsection}{0pt}{10pt}{2pt}
\titleformat{\subsubsection}[runin]{\normalfont\bfseries}{}{0pt}{}
\titlespacing*{\subsubsection}{0pt}{10pt}{2pt}

\usepackage{ccaption}
\hangcaption
\captiondelim{. }
\captionstyle{\small\raggedright}

\usepackage{natbib}
\bibpunct{(}{)}{;}{a}{,}{,}

\begin{document}

\twocolumn[
\begin{@twocolumnfalse}
% Do not edit this section ----
\hfill Proceedings of the RAAD 2013 \\ \null
\hfill 22nd International Workshop on Robotics in Alpe-Adria-Danube Region\\ \null
\hfill September 11-13, 2013, Portoro\v{z}, Slovenia\\ \null
\vspace{20pt}
% -----------------------------

\begin{center}
{\fontsize{14}{14}\selectfont \bfseries
% INSERT YOUR TITLE HERE ------
Semantic Image Search for Robotic Applications
}
% -----------------------------
\end{center}
\begin{adjustwidth}{1.5cm}{1.5cm}
\begin{center}
\begin{large}
% EDIT AUTHORS' NAMES HERE ----
Tomas Kulvicius\textsuperscript{a}, Irene Markelic\textsuperscript{a}, Minija Tamosiunaite\textsuperscript{a} and Florentin W\"org\"otter\textsuperscript{a}
% -----------------------------
\end{large}
\\ \vspace{10pt}
\textsuperscript{a}\hskip -3pt \textit{
% EDIT THE FIRST AFFILIATION DESCRIPTION
Georg-August-Universit\"at G\"ottingen \\
Bernstein Center for Computational Neuroscience \\
Department for Computational Neuroscience \\
III Physikalisches Institut - Biophysik \\
Friedrich-Hund Platz 1, DE-37077 G{\"o}ttingen, Germany \\
E-mail: \{tomas,irene,minija,worgott\}@physik3.gwdg.de }% --------------------------------------
\\ \vspace{10pt}
%\textsuperscript{b}\hskip -3pt \textit{
% EDIT THE SECOND AFFILIATION DESCRIPTION
%Department \\
%University, Country \\
%E-mail: john.doe@uni.edu}
% ---------------------------------------
\end{center}
\end{adjustwidth}
\vspace{16pt}
\begin{adjustwidth}{2cm}{2cm}
{\fontsize{9}{9}\selectfont {\bfseries Abstract.}
% INSERT YOUR ABSTRACT HERE ----
Generalization in robotics is one of the most important problems. New generalization approaches use internet databases in order to solve new tasks. Modern search engines can return a large amount of information according to a query within milliseconds. However, not all of the returned information is task relevant, partly due to the problem of polysemes. Here we specifically address the problem of object generalization by using image search. We suggest a bi-modal solution, combining visual and textual information, based on the observation that humans use additional linguistic cues to demarcate intended word meaning. We evaluate the quality of our approach by comparing it to human labelled data and find that, on average, our approach leads to improved results in comparison to Google searches, and that it can treat the problem of polysemes.

% ------------------------------
}
\end{adjustwidth}
\vspace{8pt}
\begin{adjustwidth}{2cm}{2cm}
{\fontsize{9}{9}\selectfont {\bfseries Keywords.}
% INSERT YOUR KEYWORDS HERE ----
Internet-based Knowledge, Polysemy, Semantic Search, Image Database Cleaning
}
\end{adjustwidth}
\vspace{30pt}
\end{@twocolumnfalse}
]

% HERE BEGINS YOUR PAPER

\section{Introduction}
\label{sec:Introduction}

Humans can generalize to new tasks very quickly whereas for robots this is still not an easy task which makes it one of the most important and relevant problems in robotics. One of the most common approaches in generalization is learning from previous experiences \citep{Ude2010,Nemec2011,Kober2012,Kronander2011}. Some new approaches use internet databases in order to generalize to new situations \citep{Tenorth2011,Beetz2011,Tamosiunaite2011}. In particular, here we are interested in generalization in object domain by using image search. Although modern search engines like Google or Yahoo do an amazing job in returning a large number of images according to a query within milliseconds, not all of the returned images are task/context-relevant. 
A reason for spurious results is that most image searches rely on text-based queries,
which is justified, since visual and textual information are dual to some degree.
An \textit{image} of a cup can be interpreted as the visual representation of the concept cup, whereas the \textit{word}
cup can be seen as a linguistic handle to the concept cup as represented in the human mind \citep{InternalRepresentationGrush2004}.
Therefore, existing tools for text-based information retrieval applied to image search can lead to relatively good results \citep{PageRank}.
Problems arise mainly due to ambiguities: 
1) The same linguistic handle can map to several, different concepts,
 e.g., homonyms and polysemes. An example
is the word ``jaguar'' which can refer to a car or an animal.
Without any further information, e.g., contextual information, it is not possible to infer which domain 
is actually referred to.
2) Text-based image search relies on the assumption that textual information that is somehow related to an image, e.g., text
placed close-by an image on a web page
refers to the image content \citep{PageRank}.
This assumption is reasonable, however not always correct, e.g., not every web-page creator names images according to their content.

A lot of effort has been spent on trying to resolve the problem of obtaining unclean image search results, often with the goal of object detection or image categorization, by
making additional use of image content in form of visual cues, e.g., features like local image patches, 
edges, texture, color, deformable shapes, etc. \citep{Fergus03,Fergus04,Fergus05,Multimodal,WebAnimals,WebHarvesting,Ayatollah,FeiFei,BuildTextFeats,GooglePaper,CelebritiesClustering}. All these approaches use textual information, too. Either implicitly by using the results of text-based image search engines
 e.g., \citet{Fergus05,Fergus04}, 
or constructing their
own image search \citep{WebHarvesting,WebAnimals,CelebritiesClustering}, or explicitly, by making use of image tags and labels as found in photo-sharing websites like Flickr
\citep{Multimodal,BuildTextFeats,WebAnimals}. 
An interesting work is \citet{BuildTextFeats}, 
because it is somewhat inverse to the standard procedure: 
Instead of using images with similar text labels to obtain image features for classification, they reverse
the problem and use similar images to obtain textual features.

To our knowledge all of the aforementioned approaches achieve an improved precision of the result set, however, none can 
automatically cope with the problem of polysemes. For example in 
\citet{Fergus04} a re-ranking of images obtained from Google searches was proposed, based on the observation that
images related to the search are visually similar while unrelated images differed. 
This ``visual consistency'', what we will here call inter-image similarity, was 
measured using a probabilistic, generative image model, and the EM-algorithm was used for estimating the model parameters from image features.
Naturally, due to the underlying assumption, this will not work well for homonyms, since 
for these many images that are actually closely related to the search can have a very different appearances.
A similar problem was faced in \citet{Fergus05}, where
an extended version of pLSA (probabilistic Latent Semantic Analysis) was used to learn a 
clustering of images obtained from a Google search.
A solution suggested in \citet{WebAnimals} copes with the polysemes problem but requires human supervision for this stage.
Google text search is used to collect webpages for 10 animals. Then 
LDA (Latent Dirichlet Allocation) is applied to text from these pages to discover a set of latent topics. 
Images extracted from the webpages are then assigned 
to the identified topics, according to their nearby word likelihood. 
The problem of polysemes is tackled by a human user who manually 
selects or rejects these image sets.

Here, we present our approach which we call SIMSEA (Semantic Image SEArch) 
which also aims at increasing the precision of Internet image search results. Its most prominent advantage is that it can cope near-to 
automatically with polysemes.
This is achieved by exploiting the fact that also humans need to resolve ambiguities in
every-day speech, e.g., we may say ``the bank - that you can \textit{sit} on'' 
to distinguish it from the bank that deals with money.
Thus, we give additional cues to demarcate our intended meaning of a word.
Here, we combine this linguistic refinement with the image-level in the following way: 
We conduct several different image searches, where we pair the basic search term with an additional linguistic cue. 
E.g., if interested in the category ``cup'', (the basic search term), we search for ``coffee cup'', ``tea cup'', etc.
The expectation is that images that are retrieved by more than one of these subsearches are more likely to be of interest, than those that are retrieved only once. Note that for simplicity, in this paper we defined additional cues manually. In general, automated extraction of object descriptors (cues) can be done using methods of natural language processing \citep{Cimiano2006,Olivie2011,McAauley2012}, however, this is out of the scope of the current paper.

To compute the similarity between images from different subsearches, we use a ``Bag-of-Words'' (or ``Bag-of-Features'') representation as often used in image classification. More precisely, we compute a codebook based on PHOW \citep{Bosch07,bosch2007a} features. However, also other, or additional,
features are possible.

In addition to this procedure for achieving cleaner search results, we propose a ranking of the retrieved images, which
is simply based on the idea that an image is the more relevant the more subsearches it, or a very similar match, is contained in.

We evaluate the quality of the obtained image set, and our proposed ranking, by comparing it to human labelled data. 

The paper is structured as follows:
We give a detailed description of our procedure in section \ref{sec:Methods}, followed by the explanation of how we evaluated our method and the presentation of the achieved results in section \ref{sec:Evaluation}. Finally we discuss and conclude our work in section \ref{sec:Discussion}.

\begin{figure*}
\centering
\includegraphics[width=0.98\textwidth]{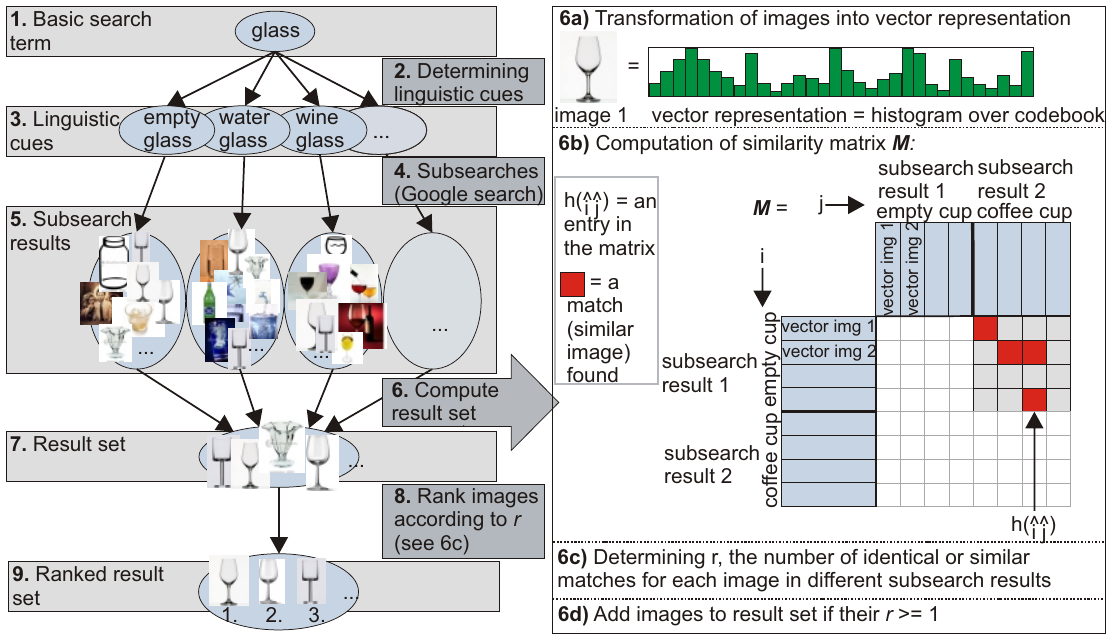}
\caption{\label{fig:MethodOverview} The different stages of SIMSEA sketched for the category ``glass''. Note that $M$, for clarity, is only shown for the first two subsearch results.}
\end{figure*}

\section{Methods}

\subsection{SIMSEA Overview}
\label{sec:Methods}
The approach is summarized in Fig. \ref{fig:MethodOverview} and an overview
on its stages, which are enumerated in the figure, are described below,
followed by a more detailed description of each stage in the paragraphs 
\ref{subsection:LinguisticCuesandSubSearches} and 
\ref{subsection:SubsetRetrieval}.

The goal is to find ``clean'' results for image searches with respect to given task/context.
For that we conduct several image searches to which we refer as \textit{subsearches} (4), see Fig. \ref{fig:MethodOverview}.
A subsearch is conducted using the \textit{basic search term} (1) with an additional \textit{linguistic cue} (2+3). 
E.g., if interested in the category ``cup'', we search for ``coffee cup'', ``tea cup'', etc. (using Google). The set of images retrieved
by a subsearch is consequently referred to as \textit{subsearch results} (5).
The expectation is that images that are retrieved by more than one subsearch
are more likely to be task/context-relevant than those that do not, they form the final \textit{result set} (6+7).
We do not consider only images that have exact copies in other subsearch result sets, but instead relax this demand and also consider images as relevant if
merely a similar image is returned by another subsearch.

Finally, we suggest to rank the retrieved result set (8).
The ranking is supposed to indicate how relevant a given image is, e.g., a glass-image with a high ranking factor
should be considered to be very likely a true representative of the category glass, whereas an image with a 
low ranking factor can be considered to be very likely not a good representative of its class.
As stated, we assume that images that
have similar counterparts in other subsearch results are
more likely to be relevant. This measurement 
can be used as a simple relevance ranking of the resulting images:
The more often an image (or a similar counterpart) occurred in 
other subsearches the higher its relevance.

\subsection{Linguistic Cues and Subsearches}
\label{subsection:LinguisticCuesandSubSearches}
We investigate four different categories (basic search terms) taken from a kitchen scenario: ``cup'', ``glass'', ``milk'' and ``apple''. 
Glass and apple are homonyms (vision, drinking, and material; or brand and fruit). 
Milk is another special case, because as a liquid it usually comes in a more or less characteristic container.
For each of the four categories we conduct a varying number of subsearches in which we combine the basic search term with an additional
linguistic cue. For the category milk we conduct six subsearches, namely: ``cold milk'', ``fresh milk'',  ``healthy milk'', ``tasty milk'', and ``hot milk''.
 We also conduct a query with the basic search term ``milk'' without any additional cue. 
The linguistic cues we use for apples are: ``delicious'', ``green'', ``red'', ``ripe'', ``sour'', ``sweet'', and ``unripe''. In addition we search for the basic terms ``apple'' and ``apples''.
For glass we use: ``empty'', ``full'', ``juice'', ``milk'', ``water'', ``wine'', and the word ``glass'' by itself. For cup: ``coffee'', ``full'', ``tea'', and simply ``cup''.
The strategy for selecting the cues was to select those that restrict the domain to the desired kitchen domain. 

\subsection{Computing the Result Set}
\label{subsection:SubsetRetrieval}

As explained, the expectation is that images that have similar counterparts (or matches) in the other subsearch results are
likely to meet the user expectations.
To be able to measure inter-image similarity we use a ``Bag-of-Words'' approach. 
In such an approach each image is represented by a histogram over a fixed number of so-called ``visual words'' 
which are also often referred to as ``codebook''. These visual words are usually created based on local image features. In our case we use 
PHOW features (which are explained below) but other features can be used, too. 

First, the codebook needs to be generated. For that we take a small, randomly chosen subset of images, we use 40, from each category.
We compute PHOW features for all these 160 (40$\times$4 categories) images which we then quantize into $k$ vectors - the visual words -
using $k$-means clustering. We set $k$ to 200.

After having created the codebook, we can represent each image by a vector, a histogram over the codebook, which is computed as follows
(step 6a in Fig. \ref{fig:MethodOverview}):
We determine the PHOW features of each image $i$, map these to the $k$ visual words that make up the codebook, and compute the histogram
which counts which visual word occurred how
many times in the given image. This histogram, called ``vector image $i$'' in Fig. \ref{fig:MethodOverview}, is then used to represent image $i$.

To compute the similarity matrix $M$ (step 6b), we consider all images $i$, $j$,
which are contained in different subsearch result sets.
For each such image pair we compute the Hellinger distance $h(\hat{i}\hat{j})$, described in detail below. Here, $\hat{i}$ and $\hat{j}$ refer to the vector representation 
of the two images. This leads to an upper diagonal matrix $M$, as indicated in Fig. \ref{fig:MethodOverview}. 

In step 6c we then determine for each image $i$ the number of similar matches or counterparts for that image which we refer to as the ranking factor $r_i$.
We consider two images to be matches if their
Hellinger distance is above a certain threshold. 
Determining $r_i$ is simply counting the number of matches for that image.
For example, the first row of matrix $M$ in Fig. \ref{fig:MethodOverview} shows all matches for the first image in the subsearch for ``empty cup'' in orange.
Since there is only one match, $r$ for this image is one. For the second image of the same subsearch there are two matches, thus $r$ for this image is two, etc.

Finally, we select those images to be part of the final result set whose ranking factor $r$ is larger than one (step 6d).

The ranking, step 8, is trivial, it merely consists of ordering the images from the computed result set according to their ranking factor.
The ranking is supposed to indicate how relevant a given image is, e.g., a glass-image with a high ranking factor
should be considered to be very likely a true representative of the category glass, whereas an image with a 
low ranking factor can be considered to be very likely not a good representative of its class.

\begin{figure}
\centering
\includegraphics[width=0.98\columnwidth]{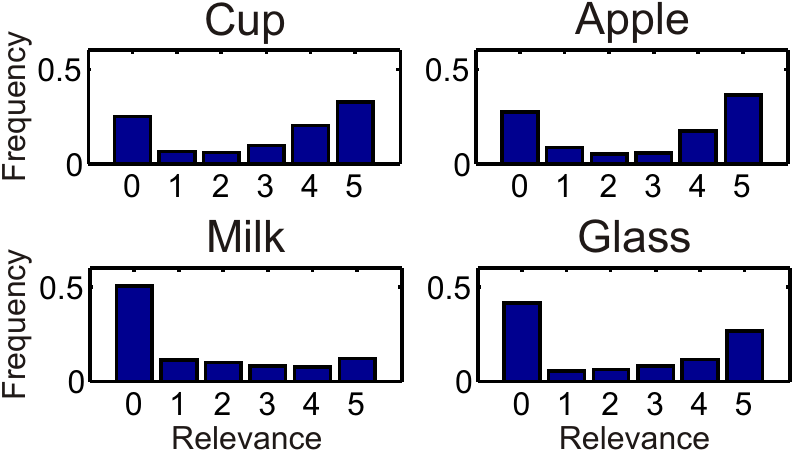}
\caption{\label{fig:RelevanceDistribution} 
Histogram of image category membership assigned by the five human subjects from which we derive the image relevance.}
\end{figure}

%\subsubsection{PHOW Features:}
%\label{subsubsection:Phow}

Pyramid Histogram of Visual Words \citep{Bosch07,bosch2007a} are  
state-of-the-art image descriptors based on a variant of dense SIFT \citep{Lowe2004}.
A grid with a self-defined spacing (here we use 5 pixels) is laid over an image and at each grid point
four SIFT descriptors, varying in radii to allow for scale variations, are computed.
This can be done on various levels, hence ``Pyramid'', but here we suffice with the first level, thus, to be precise we are actually using HOW descriptors.
We use the VLFeat library \citep{vedaldi10vlfeat} to compute the PHOW descriptors and the subsequent vector representation of the images.

%\subsubsection{Similarity:}
%\label{subsubsection:Similarity}

To compute the similarity between image pairs we use the Hellinger distance\footnote{We also used the $\chi^2$-distance, which gave very similar results. The Hellinger distance has the advantage to be bounded.}. 
The Hellinger distance between two distributions $P$ and $Q$ is denoted $H(P,Q)$ and satisfies 
    $0\le H(P,Q) \le 1$ (where $1$ denotes large distances and $0$ no distances, i.e., identical images). It is defined as follows.
\begin{equation}
     H(P,Q) = \sqrt{1 - BC(P,Q)},
\end{equation}
where $BC$ denotes the Bhattacharyya coefficient which, in the discrete case, is defined as:
\begin{equation}
    BC(P,Q) = \sum_{x\in X} \sqrt{P(x) Q(x)}.
\end{equation}
Here $X$ denotes the common domain over which the two distributions are defined.
We define two images to be similar if their Hellinger distance is above a fixed threshold (we use 0.15, experimentally chosen).
For $n$ subsearches, where $s_i$ denotes the $i$'th subsearch, we compare each image from each subsearch to all images from all other subsearches
except to its own. Thus, if the total number of images is $N=|s_1| + |s_2| + \ldots + |s_n|$ (where the vertical
bars denote the number of elements in the set), we have $C={N \choose 2} - \sum_{i=1}^n {{|s_i|}\choose {2}}$, where $C$ is the total number of
comparisons that need to be computed. This is depicted in the visualization of $M$ in Fig. \ref{fig:MethodOverview}. Note, we do not compare images from the same subsearch to each other. This is because we are not interested
in intra-subsearch similarity due to the following reason: We may receive many images of the same topic during one search but
which are unrelated to what we are interested in. If we counted the intra-subsearch similarity these images would be evaluated as highly relevant
to our search interest which they are not.

\begin{figure*}
\centering
\includegraphics[width=0.98\textwidth]{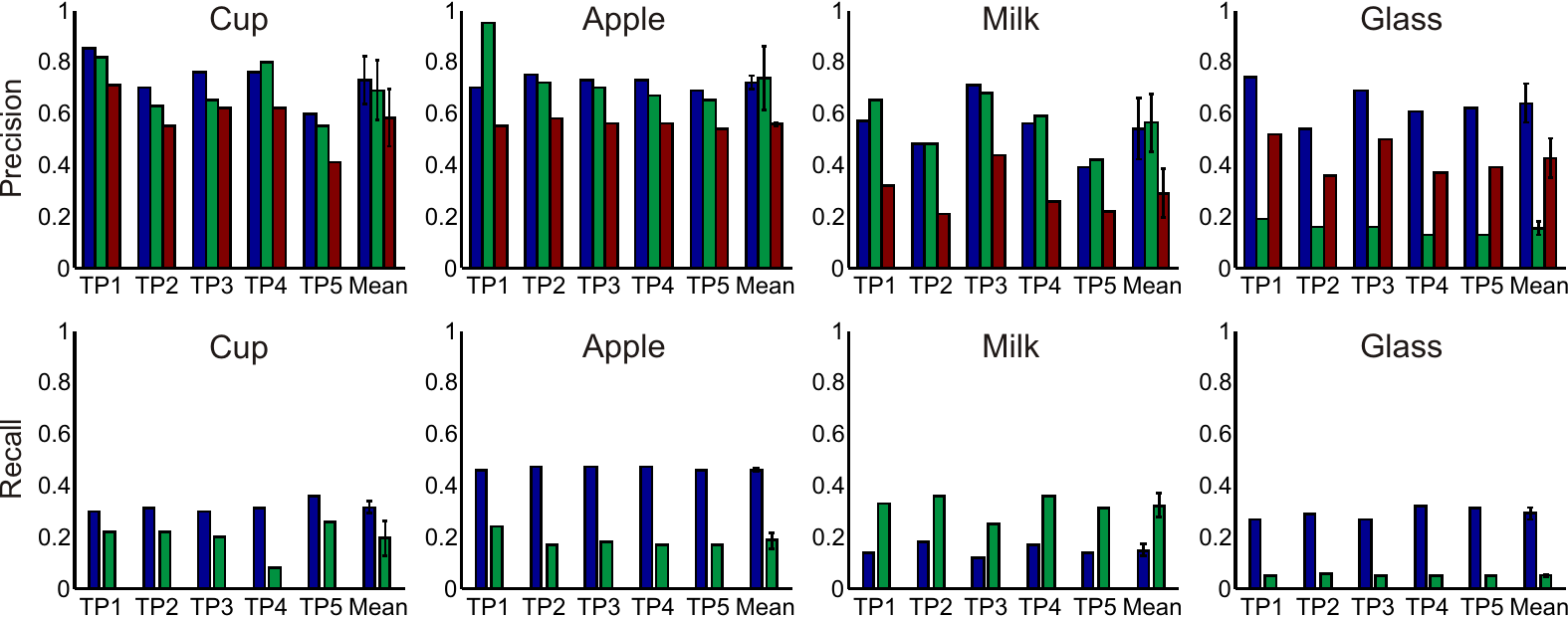}
\caption{Precision and recall of SIMSEA (blue), a standard Google search (Google, in green) and the cumulative data from all subsearches for a given category (SumGoogle, in red)
with respect to the data obtained from each test person (TP1-5) for the categories. The vertical errorbar for the mean indicates the variance.
} 
\label{fig:BarPlots}
\end{figure*}

\section{Results}
\label{sec:Evaluation}

%(edit: this part could be merged into the method section:)
Since the goal is to find a subset of images which meets the semantic expectation of the user,
we need some ``ground truth'', i.e., a set of true samples, to evaluate our algorithm.
For this issue we let several human subjects classify the same data that was input to the algorithm
according to the given categories.
This way we can gather various subjective human opinions and determine those images that get assigned the same labels by all subjects and also those
where opinions differed. 
In the following we describe the ground truth retrieval procedure.%we determine their common subset of images that belong to a category and those that do not.
%We then compare the result of the algorithm to the result produced by of each human subject and also
%to the common subset of each human.

\subsection{Ground-Truth Retrieval}

We asked five human subjects to aid in retrieving the ground-truth data to which we compare our algorithm.
Each human was instructed to decide for each image from the subsearches for milk (hot, tasty, cold, ..) if it
belonged, in his or her opinion, to the category milk. The same was done for the three other categories, cup, glass and apples. To make this evaluation as fair as possible, all humans were given precisely the same information by means of an instruction. Basically, the subjects were told that there are four categories and that they are from a kitchen scenario, thus, glass was supposed to be for drinking, and not for aiding vision, etc. 

%\subsubsection{Relevance:}
%\label{subsubsec:Relevance}
As explained, we suggest a ranking procedure which is supposed to reflect the relevance of a given image. So far we loosely defined
relevance as ``goodness of an image as class representative''. 
To evaluate our ranking result, we again require some ``ground truth'' data that we can compare it to, which we obtain as follows: 
In Fig. \ref{fig:RelevanceDistribution} 
we show a histogram indicating for each category how many of the test persons considered a given image as being member of a  
category. Since there were five test persons each image can be selected as category member between zero and five times.
We assume that images which were considered by none of the test persons as category member should be assigned
the lowest relevance, and vice versa, images considered by all test persons should be assigned 
the highest relevance. Thus, for each image we can compute a measure based on the five human subjects decisions to which we refer to as \textit{relevance} whereas the equivalent measure of SIMSEA is called \textit{ranking}. We will use these measurements in Section \ref{subsection:Results}.

\begin{figure*}
\centering
\includegraphics[width=\textwidth]{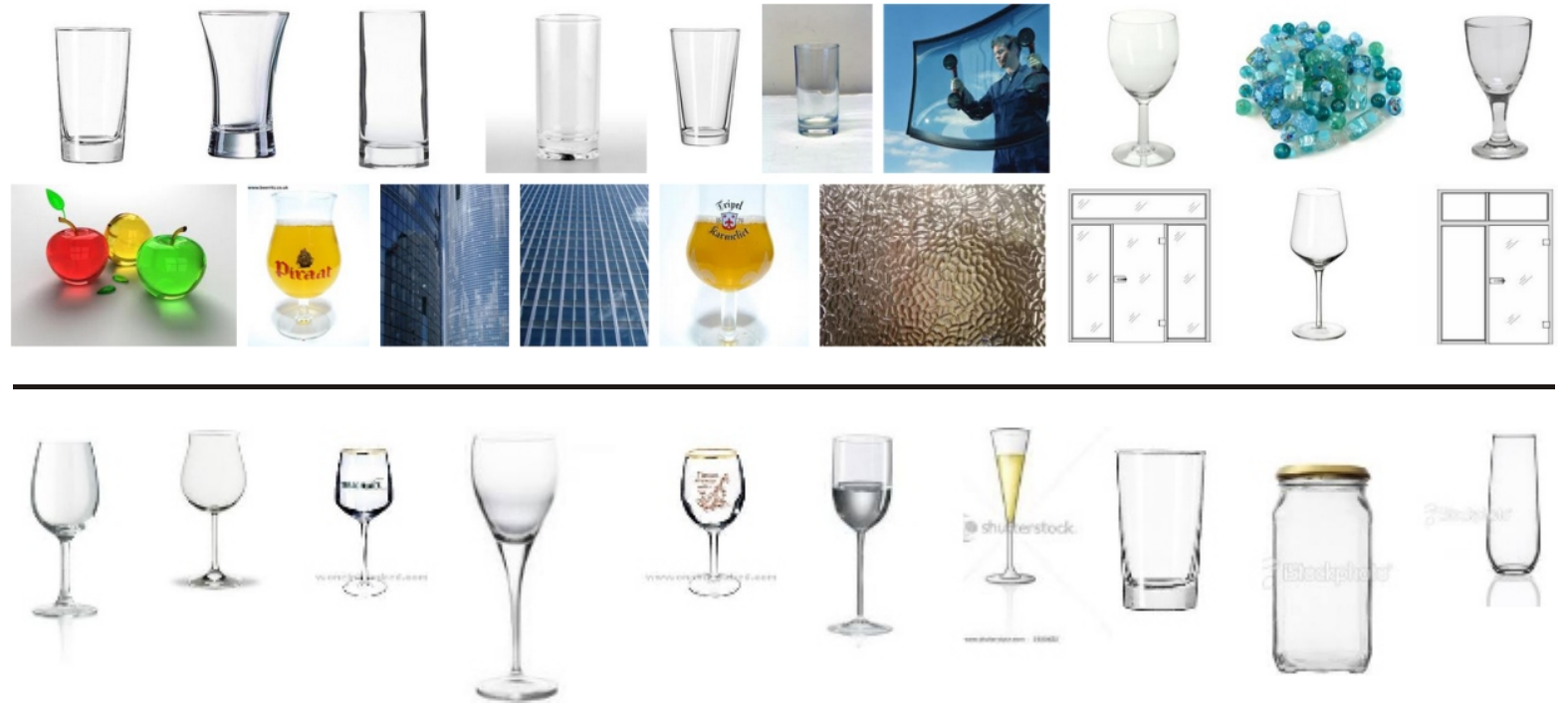}
\caption{\label{fig:BestWorstGlasses} The first 19 glasses from a Google search for the word ``glass'' (upper panel) and the best ten glasses according to the ranking obtain by using SIMSEA algorithm (bottom panel).
}
\end{figure*}

\subsection{Evaluation}
\label{subsection:Results}

We assess the quality of the algorithm by computing precision and recall on its output, see Eq. \ref{eq:precisionAndRecall},
with respect to the ground truth data from each human subject.
\begin{equation}
   \begin{split}
   \mbox{precision:=}  & (A\cap B) / |A| \\ 
   \mbox{recall:=}     & (A\cap B) / |B|,  
   \end{split}
   \label{eq:precisionAndRecall}
\end{equation}
where $A$ is the set of retrieved samples and $B$ is the set of true samples, i.e., in our case $A$ is the set of samples retrieved by SIMSEA and
$B$ is the set of samples belonging to a given category selected by each human subject. Since there were five human subjects, there are 
five true sample sets, with respect to which we compute precision and recall. The results are given in Fig. \ref{fig:BarPlots}.
We compare these results to (i) standard Google searches and also to (ii) the union of all subsearches of a given category.
For (i), we conduct standard Google searches with the basic search terms for each category, e.g., for the category milk, the set $A$ is the set of images returned
by a Google search using the search term ``milk'', and again compute precision and recall
the way described before. In
Fig. \ref{fig:BarPlots}
we refer to this evaluation as ``Google''.
For (ii) we set $A$ to the union of the images from all subsearches of a given category, i.e., if we conducted $n$ subsearches for the category
milk, we have $s_1$, $s_2$\dots$s_n$ subsearch result sets and we set $A = s_1 \cup s_2\dots \cup s_n$. In 
Fig. \ref{fig:BarPlots}
we refer 
to this evaluation as ``SumGoogle''.
Note, that for SumGoogle the recall is always one. This is because the ground truth set from all human subjects is a subset 
of the union of subsearches for a category, in other words $B\subset A$.

To be useful, precision and recall of SIMSEA should be higher than those of the standard Google search and SumGoogle. 
In other words, most human subjects should find that the output of SIMSEA gives more relevant results than the Google standard search and SumGoogle (precision), and also that
SIMSEA returns more of the overall available relevant samples (recall).
It can be seen from 
Fig. \ref{fig:BarPlots}
that except for the category ``milk'' SIMSEA indeed outperforms the standard Google search and SumGoogle.

For the category ``cup'', test person 4 (TP4) agrees more with the results of the Google search, but all other human subjects consider more
images retrieved by the automatic routine to be important. It can also be seen that the values for precision and recall differ between the subjects
which shows, what we had already expected, that assigning images to a certain category also depends on subjective opinions.
For the category ``apple'', TP1 shows a very clear preference for the Google search results. Due to TP1 also the precision is higher
for the Google search than the automatic routine.
However, TP1's opinion is not in accordance with the that of
the other subjects, which all have a precision value around 0.7 and therefore we consider this to be an outlier. Without TP1's influence 
SIMSEA outperforms the Google search for ``apple'', too.

For the category ``milk'' we can observe a different case, most human subjects are more in accordance with the results of the Google standard search.
A possible reason for that can be found in Fig. \ref{fig:RelevanceDistribution}.
We see that for all categories there are clear peaks for images that all human subjects consider as category member and for those that
all human subjects consider to not be category members. Except for the category ``milk''. Here, the relation between images with
full voting or relevance (all five human subjects) and ambiguous decisions, i.e., where for example only two out of three subjects considered an
image as relevant, is higher than compared to the other categories. In other words, there are many images in the milk category for which even humans find a clear decision difficult.
This might be due to the fact that we have already stated that milk as a liquid is depicted to be contained in more or less characteristic containers. 
We can assume that for this reason SIMSEA is not performing well for this category either. 

To select ``clean'' images from a Google search we use a ranking $r$ as described above, i.e., frequency
with which they occur in the different subsearches. To visualize the effect of the ranking we show the best ten glasses in Fig. \ref{fig:BestWorstGlasses}, bottom panel. In the upper panel we show the first 19 images returned by a Google search for the word ``glass''. We can see that Google search results include images of glasses from domains others then the desired kitchen domain (in this case $\approx 42 \%$). SIMSEA in contrast was successful in eliminating those.

\section{Discussion}
\label{sec:Discussion}

We proposed a method based on the combination of linguistic cues with the image domain
that is useful for retrieving cleaner results in image searches, in particular it is able to tackle the
problem of polysemes.
This is a novel approach and we have given the proof of principle by showing that it indeed leads to cleaner search results.

In addition we suggested a ranking, based on the occurrence frequency of images between different subsearches.
We could show that this roughly reflects a human based relevance measure.

Although we have introduced the notion of linguistic cues, we have not tackled the issue where these cues
might come from, or how they should best be chosen. Automated extraction of object descriptors (cues) can be done using methods of natural language processing \citep{Cimiano2006,Olivie2011,McAauley2012}. However, this is an issue falling in the domain of linguistics and is not the core of this paper.

It is obvious that our method can only be as good as the subsearch results which depend on the ``right'' linguistic cues. If unrelated images occur in many of the subsearches, these images will erroneously be part of the result set. 

Similar to the effectiveness of human linguistic refinement to distinguish intended meaning from other, our method has its strength when dealing with polysemes or homonyms. For example, the result for the category ``glass'' is very good, where it had to distinguish drinking glasses from the material glass and the vision aid. In contrast the method did not perform well for the liquid ``milk''.

Another critical issue is the similarity function between images. Here we used PHOW features and the Hellinger distance, which works satisfactory, but also sometimes lead to artefacts, i.e., images that do not appear similar to humans can sometimes turn out to be very similar when using PHOW features. Here, different features and metrics may lead to an improvement of the method. Another option can be to follow the idea of \citep{CelebritiesClustering} and to learn an appropriate metric by solving a constrained optimization problem.

In summary, we believe that this a novel and promising idea for data ``cleaning'' which can be used to automatically form training data sets using Internet search which later can be used for object classification/recognition and generalization. In future work we are going to include more classes and make such image search completely automatic by augmenting it with an automated extraction of object descriptors from language.

\section{Acknowledgements}

The research leading to these results has received funding from the European Community's Seventh Framework Programme FP7/2007-2013 (Programme and Theme: ICT-2011.2.1, Cognitive Systems and Robotics) under grant agreement no. 600578, ACAT.

\section{References}
\bibliographystyle{apalike}
%{\small
%\bibliography{LingoPaperReferences}

\begin{thebibliography}{}

\bibitem[A.D.~Holub, 2008]{CelebritiesClustering}
A.D.~Holub, P.~Moreels, P.~P. (2008).
\newblock Unsupervised clustering for google searches of celebrity images.
\newblock {\em 8th IEEE Int. Conf. Automatic Face and Gesture Recognition}.

\bibitem[Beetz et~al., 2011]{Beetz2011}
Beetz, M., Klank, U., Kresse, I., Maldonado, A., M\"osenlechner, L., Pangercic,
  D., R\"uhr, T., and Tenorth, M. (2011).
\newblock {Robotic Roommates Making Pancakes}.
\newblock In {\em 11th IEEE-RAS Int. Conf. on Humanoid Robots}, pages 529--536,
  Bled, Slovenia.

\bibitem[Berg and Forsyth, 2006]{WebAnimals}
Berg, T.~L. and Forsyth, D.~A. (2006).
\newblock Animals on the web.
\newblock In {\em CVPR}, pages 1463--1470.

\bibitem[Bosch et~al., 2007a]{Bosch07}
Bosch, A., Zisserman, A., and Munoz, X. (2007a).
\newblock Representing shape with a spatial pyramid kernel.
\newblock In {\em ACM Int. Conf. Image and Video Retrieval}.

\bibitem[Bosch et~al., 2007b]{bosch2007a}
Bosch, A., Zisserman, A., and Muñoz, X. (2007b).
\newblock Image classification using random forests and ferns.
\newblock In {\em ICCV}, pages 1--8.

\bibitem[Brin and Page, 1998]{PageRank}
Brin, S. and Page, L. (1998).
\newblock The anatomy of a large-scale hypertextual web search engine.
\newblock {\em Comput. Netw. ISDN Syst.}, 30:107--117.

\bibitem[Cimiano, 2006]{Cimiano2006}
Cimiano, P. (2006).
\newblock {\em Ontology Learning and Population from Text: Algorithms,
  Evaluation and Applications}.
\newblock Springer Verlag.

\bibitem[Fergus et~al., 2005]{Fergus05}
Fergus, R., Fei-Fei, L., Perona, P., and Zisserman, A. (2005).
\newblock Learning object categories from google's image search.
\newblock In {\em 10th IEEE Int. Conf. Computer Vision}, volume~2, pages
  1816--1823.

\bibitem[Fergus et~al., 2003]{Fergus03}
Fergus, R., Perona, P., and Zisserman, A. (2003).
\newblock Object class recognition by unsupervised scale-invariant learning.
\newblock In {\em CVPR}, pages 264--271.

\bibitem[Fergus et~al., 2004]{Fergus04}
Fergus, R., Perona, P., and Zisserman, A. (2004).
\newblock A visual category filter for google images.
\newblock In {\em 8th Europ. Conf. Computer Vision}, pages 242--256.

\bibitem[Grush, 2004]{InternalRepresentationGrush2004}
Grush, R. (2004).
\newblock The emulation theory of representation: Motor control, imagery, and
  perception.
\newblock {\em Behavioral and Brain Sciences}, 27:377–442.

\bibitem[Guillaumin et~al., 2010]{Multimodal}
Guillaumin, M., Verbeek, J., and Schmid, C. (2010).
\newblock Multimodal semi-supervised learning for image classification.
\newblock In {\em CVPR}.

\bibitem[jia Li et~al., 2007]{FeiFei}
jia Li, L., Wang, G., and Fei-fei, L. (2007).
\newblock Optimol: automatic online picture collection via incremental model
  learning.
\newblock In {\em CVPR}.

\bibitem[Jing and Baluja, 2008]{GooglePaper}
Jing, Y. and Baluja, S. (2008).
\newblock Visualrank: Applying pagerank to large-scale image search.
\newblock {\em Pattern Analysis and Machine Intelligence, IEEE Transactions
  on}, 30(11):1877 --1890.

\bibitem[Khan et~al., 2011]{Ayatollah}
Khan, I., Roth, P.~M., and Bischof, H. (2011).
\newblock Learning object detectors from weakly-labeled internet images.
\newblock In {\em 35th OAGM/AAPR Workshop}.

\bibitem[Kober et~al., 2012]{Kober2012}
Kober, J., Wilhelm, A., Oztop, E., and Peters, J. (2012).
\newblock Reinforcement learning to adjust parametrized motor primitives to new
  situations.
\newblock {\em Auton. Robots}, 33(4):361--379.

\bibitem[Kronander et~al., 2011]{Kronander2011}
Kronander, K., Khansari-Zadeh, M., and Billard, A. (2011).
\newblock Learning to control planar hitting motions in a minigolf-like task.
\newblock In {\em 2011 IEEE/RSJ Int. Conf. Intelligent Robots and Systems},
  pages 710 --717.

\bibitem[Lowe, 2004]{Lowe2004}
Lowe, D.~G. (2004).
\newblock Distinctive image features from scale-invariant keypoints.
\newblock {\em Int. J. Comput. Vision}, 60:91--110.

\bibitem[McAuley et~al., 2012]{McAauley2012}
McAuley, J.~J., Leskovec, J., and Jurafsky, D. (2012).
\newblock Learning attitudes and attributes from multi-aspect reviews.
\newblock In {\em International Conference on Data Mining}.

\bibitem[Nemec et~al., 2011]{Nemec2011}
Nemec, B., Vuga, R., and Ude, A. (2011).
\newblock Exploiting previous experience to constrain robot sensorimotor
  learning.
\newblock In {\em 11th IEEE-RAS Int. Conf. Humanoid Robots}, pages 727--732.

\bibitem[Olivie et~al., 2011]{Olivie2011}
Olivie, J., Christianson, C., and McCarry, J. (2011).
\newblock {\em Handbook of natural Language Processing and Machine
  Translation}.
\newblock Springer.

\bibitem[Schroff et~al., 2007]{WebHarvesting}
Schroff, F., Criminisi, A., and Zisserman, A. (2007).
\newblock Harvesting image databases from the web.
\newblock In {\em 11th IEEE Int. Conf. on Computer Vision}, pages 1 --8.

\bibitem[Tamosiunaite et~al., 2011]{Tamosiunaite2011}
Tamosiunaite, M., Markelic, I., Kulvicius, T., and Worgotter, F. (2011).
\newblock Generalizing objects by analyzing language.
\newblock In {\em 11th IEEE-RAS Int. Conf. Humanoid Robots}, pages 557--563.

\bibitem[Tenorth et~al., 2011]{Tenorth2011}
Tenorth, M., Klank, U., Pangercic, D., and Beetz, M. (2011).
\newblock {Web-enabled Robots -- Robots that Use the Web as an Information
  Resource}.
\newblock {\em Rob. \& Automat. Magazine}, 18(2):58--68.

\bibitem[Ude et~al., 2010]{Ude2010}
Ude, A., Gams, A., Asfour, T., and Morimoto, J. (2010).
\newblock Task-specific generalization of discrete and periodic dynamic
  movement primitives.
\newblock {\em IEEE Trans. Rob.}, 26(5):800--815.

\bibitem[Vedaldi and Fulkerson, 2010]{vedaldi10vlfeat}
Vedaldi, A. and Fulkerson, B. (2010).
\newblock Vlfeat -- an open and portable library of computer vision algorithms.
\newblock In {\em 18th annual {ACM} Int. Conf. Multimedia}.

\bibitem[Wang et~al., 2009]{BuildTextFeats}
Wang, G., Hoiem, D., and Forsyth, D. (2009).
\newblock Building text features for object image classification.
\newblock In {\em IEEE Int. Conf. Computer Vision and Pattern Recognition},
  pages 1367 --1374.

\end{thebibliography}
%}

\balance

\end{document}